\documentstyle[twoside,fleqn,espcrc2,epsfig]{article}

\newcommand{\eq}{\begin{equation}}
\newcommand{\en}{\end{equation}}
\newcommand{\eqa}{\begin{eqnarray}}
\newcommand{\ena}{\end{eqnarray}}
\newcommand{\be}{\begin{equation}}
\newcommand{\ee}{\end{equation}}
\newcommand{\ba}{\begin{eqnarray}}
\newcommand{\ea}{\end{eqnarray}}

\newcommand{\ZZ}{\hbox{{\rm Z{\hbox to 3pt{\hss\rm Z}}}}}
\newcommand{\Zt}{\ZZ_2}

\newcommand{\AmS}{{\protect\the\textfont2
  A\kern-.1667em\lower.5ex\hbox{M}\kern-.125emS}}

\title{Non-perturbative states in the three-dimensional $\phi^4$ theory}
\author{M. Caselle
 \address{Dipartimento di Fisica Teorica, Universit\'a di Torino and 
INFN, sezione di Torino,\\
via P.Giuria 1, I--10125 Torino, Italy},
M. Hasenbusch
\address{Humboldt Universit\"at zu Berlin, Institut f\"ur Physik\\
Invalidenstr. 110, D-10115 Berlin, Germany}
and P. Provero\address{Dipartimento di Scienze e Tecnologie Avanzate\\
Universit\`a del Piemonte Orientale, I--15100 Alessandria, Italy}
\thanks{presented by P. Provero}}
       
\begin{document}

\begin{abstract}
We study the spectrum of massive excitations of the three-dimensional
$\phi^4$ and Ising models, in the broken-symmetry phase. Using a 
variational method, we show that the spectrum contains all the $0^+$
states that one expects from duality with the glueball spectrum of the
$\Zt$ gauge model. From the point of view of continuum $\phi^4$ theory,
we show that at least one of the states we find has a non-perturbative origin.
\end{abstract}
\maketitle
\vskip-0.5cm
\section{DUALITY AND UNIVERSALITY}
The three-dimensional Ising model is related to two other important
three-dimensional theories: by duality, to the $\Zt$ gauge model, and
by universality, to $\phi^4$ theory. 
\par
Duality is an exact
equality of partition functions: it means that the Ising model and the
$\Zt$ gauge model are different descriptions of the same physics. 
In particular, the broken symmetry phase of the Ising model is 
equivalent to the confined phase of the gauge model. 
\par
Universality tells us that the Ising model and $\phi^4$ theory behave
in the same way when approaching the critical point. Universal
quantities are the same in a critical region around the transition
point. Indeed, the universal quantities of the Ising universality
class have been predicted to great accuracy using {\em perturbative}
$\phi^4$ theory (see {\it e.g.} Ref.~\cite{zinn-justin}).
\par
Let us apply the tools of universality and duality to the problem of
determining the spectrum of massive excitations of the $3D$ Ising model
in the broken symmetry phase. The glueball spectrum in the $Z_2$ gauge
model has been thoroughly studied numerically: it turns out to be
a rich spectrum with many excitations in various angular momentum 
channels. 
On the other hand, we certainly do not expect to find an 
interesting spectrum in perturbative $ \phi^4$ theory, which
describes just one particle. 
\par
Therefore it seems that duality and universality lead to contradictory
expectations about the spectrum of the Ising model and 
$\phi^4$ theory in the broken phase.
This work clarifies these issues by a numerical evaluation of the spectrum
of both models, performed with a new variational procedure.
\vskip-0.5cm
\section{MONTE CARLO DETERMINATION OF THE SPECTRUM}
The spectrum of a model is extracted from Monte Carlo simulations by
studying the long distance exponential decay of correlation functions.
It is convenient to study time slice observables: for example if $\phi$
is the order parameter one defines
\be
S(t)=\frac{1}{L^2}\sum_{x,y}\phi(t,x,y)
\label{order_slice}
\ee
where $L$ is the lattice size in the $x$ and $y$ directions.
One then studies connected correlators of the $S$ operator, the advantage 
being that they behave as a pure exponential in the long distance limit:
\be
\langle S(0) S(t)\rangle_c\sim\; c e^{-m|t|}\ \ .
\ee 
\noindent
For a theory with a non-trivial spectrum one expects to see subleading
exponentials as well:
\be
\langle S(0) S(t)\rangle_c\sim c_1 e^{-m_1|t|}+ c_2 e^{-m_2|t|}+\dots
\label{non_trivial}
\ee
Therefore a procedure to determine the spectrum from Monte Carlo
simulation is to measure time slice correlations and fit them to 
Eq.~(\ref{non_trivial}).
\par
However there is a more effective method, inspired by what is 
commonly done in lattice gauge theory to determine the glueball 
spectrum \cite{glueballs}. One introduces a basis of 
suitably defined (time slice) operators $\hat O_i(t)$
and then computes the matrix of cross-correlators
\be
C_{ij}(t)=\langle  \hat O_i(0)\hat O_j(t)\rangle_c
\ee
$C_{ij}(t)$ is then diagonalized to read off the spectrum. 
The crucial point is of course the choice of the operator basis, which
must be carefully fine-tuned to obtain an efficient determination of
the spectrum. Our choice is described in detail in Ref.~\cite{chp}.
Here we just mention that we included the standard magnetization
Eq.~(\ref{order_slice}) in the basis $\{\hat O_i\}$ and that the 
other operators are 
defined on different length 
scales. 
\par
We simulated both the Ising model and the lattice regularized $\phi^4$
theory in the broken symmetry phase, at various temperatures 
well inside the scaling region
where universality is expected to hold. We considered the $0^+$ channel
only. It turns out that three states can be identified 
in this channel. {\it Scaling} is
perfectly statisfied since the ratios between the masses of the three
states do not change with the temperature within the scaling region.
{\it Universality} is satisfied as well since we obtain compatible
results for the ratios from the Ising and the $\phi^4$ simulations.
Therefore we can quote a single result for each mass ratio:
\ba
\frac{m_2}{m_1}&=&1.83(3)\\
\frac{m_3}{m_1}&=&2.45(10)
\ea
Also {\it duality} is satisfied since in the $\Zt$ gauge model one obtains 
mass ratios of $1.88(2)$ and $2.59(4)$ 
in the $0^+$ channel of the glueball spectrum \cite{acch}.
\par
Note that the first excited state lies below the pair production threshold:
this means that, in terms of continuum $\phi^4$ theory, it cannot be of
perturbative origin. On the other hand the state at $2.45$ times the 
fundamental mass could well be a signature of the cut in the Fourier
transform of the propagator induced by self interaction effects, and
therefore a perturbative effect.
\par
Let us analyze in more detail what we expect from perturbative $\phi^4$
theory in this respect: a simple one-loop computation (see Ref.~\cite{cut}) 
shows that the cut in the momentum space propagator implies for the
time slice correlators the behavior
\be
\langle S(0) S(t)\rangle_c\sim c_1 e^{-m_1|t|}+ c_2 \frac{e^{-2m_1|t|}}{t}
\ee
However the second term can be shown to be numerically indistinguishable
from an exponential decay with mass $\sim 2.4\; m_1$. Therefore we conclude
that while the state $m_3$ could be explained as a perturbative
self-interaction effect, the state $m_2$ is certainly of non-perturbative 
origin. 
\vskip-0.5cm
\section{BACK TO THE SPIN-SPIN CORRELATOR}
In this section we apply the knowledge we have gained of the spectrum
of the theory to the analysis of the (time slice) spin-spin correlator
\be
G(t)=\langle S(0) S(t)\rangle_c
\ee
It is useful to define the effective correlation length
\be
1/\xi_{eff}(t)=\log G(t)-\log G(t+1)
\ee
so that for $t\to\infty$ $\xi_{eff}\to 1/m$, $m$ being the fundamental mass. 
Clearly if $G(t)$ had
a purely exponential behavior then $\xi_{eff}(t)$ would be constant: 
the preasymptotic behavior of $\xi_{eff}(t)$ depends on higher mass states 
and/or interaction effects. These effects are shown in Fig.1, where
$m\;\xi_{eff}$ is shown for various values of the temperature in the Ising
model. The data from the various temperatures are perfectly compatible
with each other, signaling that the preasymptotic behavior of $\xi_{eff}(t)$
is a physical, scaling effect and not a lattice artifact. 
\begin{center}
\begin{figure}[htb]
\mbox{\epsfig{file=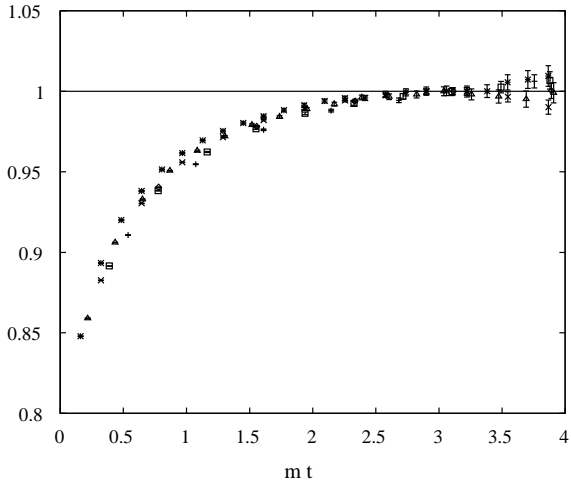}}
\vskip-0.6cm
\caption{$m\xi_{eff}(t)$ for five different values of $\beta$ for the 
Ising model}
\end{figure}
\end{center}
\par
In  Fig. 2 we use our knowledge of the spectrum to 
describe the behavior of $\xi_{eff}(t)$: the Monte Carlo
are compared to the perturbative prediction
\be
\xi_{eff}(t)=c_1\left[e^{-m_1 t}+f_{cut}(t)\right]
\label{perturbative}
\ee
(dotted line), and to the curve
\be
\xi_{eff}(t)=c_1\left[e^{-m_1 t}+f_{cut}(t)\right]+c_2 e^{-m_2 t}
\label{nonperturbative}
\ee
(solid line)
where the constatnts $m_1$, $m_2$, $c_1$, $c_2$ are taken from our
variational evaluation of the spectrum, and $f_{cut}$ is 
taken from one-loop perturbative calculations in the continuum theory
(see Ref.~\cite{cut}). The good agreement between this curve and the data
suggests that in fact the third mass $m_3$ is not a new state but a 
perturbative interaction effect.
\begin{center}
\begin{figure}[htb]
\mbox{\epsfig{file=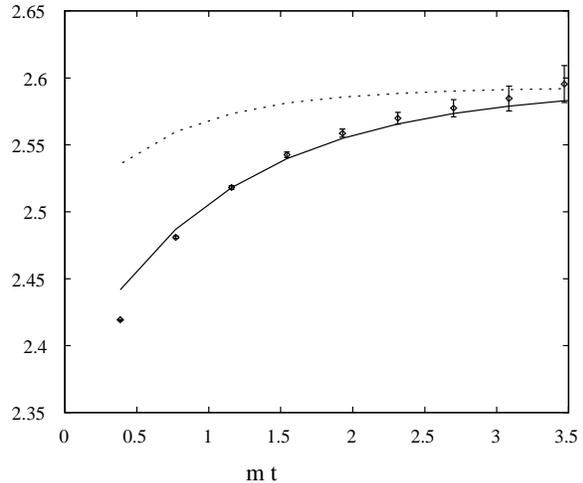}}
\vskip-0.6cm
\caption{Monte Carlo data for $\xi_{eff}(t)$ compared to 
Eqs.~(\ref{perturbative}) (dotted line) and (\ref{nonperturbative})
(solid line).}
\end{figure}
\end{center}
\vskip-0.5cm
\section{CONCLUSIONS}
The main result of our analysis is that $3D$ $\phi^4$ theory has a rich
spectrum of massive excitations that signals the existence of non-perturbative
physics. This spectrum matches accurately the corresponding spectrum of
the $3D$ Ising model, to which $\phi^4$ is related by universality, and the
glueball spectrum of the $3D$ $\Zt$ gauge model, related by duality
to the Ising model.
\par
We are currently investigating higher spin excited states,
corresponding to higher spin glueballs, and the effect of non-perturbative 
physics on the field theoretic prediction of universal quantities. 
Another interesting development would be to
investigate the same issues in other $3D$ universality classes, in 
particular in $N$-component $\phi^4$ theory.

\end{document}